\documentstyle[preprint,revtex]{aps}                  %%%To be commented
\def\gtap{\ \raisebox{-.5ex}{\rlap{$\sim$}} \raisebox{.4ex}{$>$}\ }

\newcommand{\be}{\begin{equation}}
\newcommand{\ee}{\end{equation}}
\newcommand{\beq}{\begin{eqnarray}}
\newcommand{\eeq}{\end{eqnarray}}

\begin{document}
\draft
\pagestyle{empty}                                      %%%To be commented
\centerline{%Version:\vday
                             \hfill   NTUTH--94--04}   %%%To be commented
\centerline{\hfill                 May 1994} %%%To be commented
\vfill
\begin{title}
Radiative Majorana Neutrino Masses
\end{title}
\vfill
\author{Wei-Shu Hou and Gwo-Guang Wong}
\begin{instit}
Department of Physics, National Taiwan University,
Taipei, Taiwan 10764, R.O.C.
\end{instit}
%\receipt{\today}
%
%\vskip -1cm
\vfill
\begin{abstract}

We present new radiative mechanisms for generating
Majorana neutrino masses, within an extension of
the standard model that successfully generates
radiative charged lepton masses, order by order,
from heavy sequential leptons.
Only the new sequential neutral lepton has a right-handed partner,
and its Majorana mass provides the seed for
Majorana neutrino mass generation.
Saturating the cosmological bound of $50$ eV with $m_{\nu_\tau}$,
we find that $m_{\nu_\mu}$ and $m_{\nu_e}$ could be at most
$10^{-2}$, and $10^{-3}$ eV, respectively.
The electron neutrino mass may vanish in the limit of
degenerate charged Higgs bosons.
Unfortunately, $\nu_e - \nu_\tau$ mixing is also
radiatively induced, and is too small
for sake of solving the solar neutrino problem
via the Mikheyev--Smirnov--Wolfenstein effect.

\end{abstract}
%\pacs{PACS numbers:
%12.15.Ff, %Quark and lepton masses and mixing
%12.60.Fr, %Extensions of electroweak Higgs sector
%14.80.Cp, %Nonstandard model Higgs bosons
%13.35.-r  %Decays of leptons
%14.60.Hi, %other charged heavy leptons
%14.60.Pq, %nu mass and mixing
%14.60.St  %Nonstandard nu's and nu_R etc.
%}
%%%%%%%%%%%%%%%%%%%%%%%%%% See PRL 69 #24. 1992 for the listings
\newpage
\narrowtext
\pagestyle{plain}

%\section{Introduction}

There
is no fundamental principle requiring the neutrinos to be massless.
However, if neutrinos are not exactly massless,
their near masslessness is a real mystery.
The so-called seesaw mechanism \cite{seesaw} was invented,
usually in the context of grand unified theories (GUT),
to give a natural framework for almost massless neutrinos.
It also provides a natural framework for explaining
the solar neutrino puzzle via the
Mikheyev-Smirnov-Wolfenstein effect \cite{msw}.
Recent neutrino counting experiments \cite{PDG} show that there
exists three and only three species of light neutrinos.
On the other hand, direct search for
new sequential leptons at $Z^0$ resonance
have been conducted and yields the limits \cite{PDG}
\begin{equation}
   m_E,\ m_N \gtap M_Z/2,
\end{equation}
where we denote new charge and neutral leptons as
$E$ and $N$, respectively,
and sequential means that
left-handed leptons come in weak doublets.
Although there is as yet no evidence for more sequential leptons,
it has been pointed out \cite{3p1} that if they are
found with weak scale masses,
the traditional seesaw mechanism
and $SO(10)$ based GUT theories could be in jeopardy,
and one may also face a serious challenge with the solar neutrino problem.
It is therefore of interest to investigate whether
very small neutrino masses can still be naturally generated
under such circumstances.
In this regard, Babu and Ma \cite{twow}
have built a radiative seesaw model in which
the three known neutrinos acquire radiative Majorana masses
through the exchange of two W bosons,
and could provide a natural explanation
of the solar neutrino problem.
Recently, we have sucessfully constructed
a $Z_8$ model \cite{z8} where the charged leptons acquire
mass radiatively order by order,
with all Yukawa couplings of order unity.
The three known neutrinos have no right-handed counterparts.
They are held strictly massless if one requires
the Majorana mass $m_R$ for the single
right-handed neutral lepton $N_R$ to vanish.
In this letter, we remove this {\it ad hoc} condition
and investigate the question of light neutrino Majorana mass.

Let us briefly review the model.
With minimal ``$3+1$" generations \cite{3p1},
there is only one right-handed neutral lepton $N_R$.
Consider a discrete $Z_8$ symmetry ($\omega^8 = 1$).
We assign both $\bar\ell_{iL} = (\bar\nu_{iL}$, $\bar e_{iL}$)
and $e_{iR}$ to transform as
$\omega^3$, $\omega^2$, $\omega^1$, $\omega^4$ for $i = 1-4$,
respectively, while $N_R$ transforms as $\omega^4$.
The scalar sector consists of three doublets,
$\Phi_0,\ \Phi_3$ and $\Phi_5$,
transforming as $1$, $\omega^3$ and $\omega^5$, respectively.
Thus, except for $E\simeq e_4$ and $N$,
only nearest-neighbor Yukawa couplings are allowed,
\begin{eqnarray}
  -\cal{L}_{\rm Y} &=&  f_{44} \bar\ell_{4L} e_{4R} \Phi_0
        +  \tilde f_{44} \bar\ell_{4L} N_{R}  \tilde\Phi_0 \nonumber \\
       &+& f_{43} \bar\ell_{4L} e_{3R} \Phi_3
        +  f_{34} \bar\ell_{3L} e_{4R} \Phi_3
        +  \tilde f_{34} \bar\ell_{3L} N_R \tilde\Phi_5   \nonumber \\
       &+& f_{32} \bar\ell_{3L} e_{2R} \Phi_5
        +  f_{23} \bar\ell_{2L} e_{3R} \Phi_5             \nonumber \\
       &+& f_{21} \bar\ell_{2L} e_{1R} \Phi_3
        +  f_{12} \bar\ell_{1L} e_{2R} \Phi_3 \ \ \ \ + H.c.
\end{eqnarray}
where $\tilde\Phi \equiv i \sigma_2 \Phi^*=( \phi^{0*}, -\phi^-)$
as usual. We assume $CP$ invariance for sake of simplicity.

If only $\langle\phi_0^0\rangle = v/\sqrt{2}$,
$E$ and $N$ become massive
and are naturally at $v$ scale if $f_{44}$, $\tilde f_{44}\sim 1$.
The lower generation leptons remain massless at this stage,
protected by the $Z_8$ symmetry.
To allow for radiative mass generation,
the $Z_8$ symmetry is {\it softly} broken down to $Z_2$
in the Higgs potential by $\Phi_3$-$\Phi_5$ mixing.
Explicitly,
\begin{eqnarray}
 V = \sum_i &        \mu^2_i & \Phi_i^\dagger \Phi_i
    +\sum_{i,j} \lambda_{ij} (\Phi_i^\dagger \Phi_i) (\Phi_j^\dagger \Phi_j)
    +\sum_{i \neq j} \eta_{ij}(\Phi_i^\dagger \Phi_j) (\Phi_j^\dagger \Phi_i)
                                                        \nonumber \\
   +\, [    & \tilde \mu^2 & \Phi^\dagger_3 \Phi_5
    + \zeta (\Phi_0^\dagger \Phi_3) (\Phi_0^\dagger \Phi_5)
                                     + H.c.],
\end{eqnarray}
where $\lambda_{ij}$ and $\eta_{ij}$ are symmetric.
Note that the $\zeta$ term is $Z_8$ invariant, while
the gauge invariant ``mass" $\tilde \mu^2$ transforms as $\omega^2$.
Since only $\mu_0^2 < 0$, while $\mu^2_3$ and $\mu^2_5 >0$,
$\phi^0_0 \rightarrow (v+H_0+i\chi_0)/\sqrt 2$, and $\phi^0_i \rightarrow
(h_i+i\chi_i)/\sqrt 2$ for $i=3,\ 5$. The quadratic part of
$V$ is
\begin{eqnarray}
  V^{(2)} &=& \lambda_{00} v^2 H^2_0
           + \sum_{i\neq 0} \left(
              {1\over 2}(\mu^2_i+\lambda_{0i}v^2+\eta_{0i}v^2)(h^2_i+\chi^2_i)
             +(\mu^2_i+\lambda_{0i}v^2)\vert\phi_i^+\vert^2 \right)
                                                             \nonumber \\
          &+& \tilde \mu^2 (h_3 h_5+\chi_3\chi_5
                           +\phi^-_3 \phi^+_5 + \phi^-_5 \phi^+_3)
           + {1\over 2} \zeta v^2(h_3 h_5-\chi_3\chi_5).
\end{eqnarray}
The standard Higgs boson $H_0$ couples only diagonally to heavy particles.
The nonstandard scalars $(\phi^\pm_3, \phi^\pm_5)$,
$(h_3, h_5)$ and $(\chi_3, \chi_5)$ mix via $\tilde\mu^2$ and $\zeta$ terms.
Rotating by
$\theta_+$, $\theta_H$ and $\theta_A$, we obtain the mass basis
$(H^+_1, H^+_2)$, $(H_{1}, H_{2})$ and $(A_{1}, A_{2})$, respectively.
It is clear that $\sin\theta_+ \rightarrow 0$, $(\theta_A, m_{A_1},
m_{A_2}) \rightarrow (-\theta_H, m_{H_1}, m_{H_2})$ as $\tilde\mu^2\to 0$,
while in the limit $\zeta \rightarrow 0$, one has
$(\theta_A, m_{A_1}, m_{A_2}) \rightarrow (+\theta_H, m_{H_1}, m_{H_2})$.
These two limits restore the two extra $U(1)$ symmetries of
the doublets $\Phi_3$ and $\Phi_5$.

Since $N_R$ has $Z_8$ charge assignment of $\omega^4$,
a Majorana mass term $m_R$ is in fact permitted.
We have previously set this term to zero \cite{z8} for sake of simplicity.
This was in part also for the reason of ``naturalness",
since cosmological considerations \cite{cosmb} suggest that
$m_R/v \ll 1$, in strong contrast to other dimensionless
parameters of the model.
If we do permit nonzero $m_R$, however, we see that it
breaks the chiral symmetry of the first three
generation of massless neutrinos. It provides
the seed for generating tiny Majorana neutrino masses through
mixing and nondegeneracy of the four real scalar fields,
as well as the two charged scalars.

Radiative mass generation for charged leptons
has already been discussed in ref. \cite{z8}.
Here, we concentrate on  neutrino mass generation.
The tau neutrino $\nu_\tau$ acquires
Majorana mass
via the one-loop diagram shown in Fig. 1,
which is very similar to the charged lepton
mass generation mechanism.
We find
\begin{eqnarray}
  (m_{\nu})_{33}  =  \left( {{\tilde f_{34}}^2
                              \over {32 \pi^2}} \right)
              & & \ \  \left[ \sin^2\theta_H\;
                                 G(m_{H_1}/m_R) \right.
                        +  \cos^2\theta_H\,
                                 G(m_{H_2}/m_R)        \nonumber \\
              & &   -  \sin^2\theta_A\
                                 G(m_{A_1}/m_R) \left.
                        -  \cos^2\theta_A\,
                                 G(m_{A_2}/m_R) \right]\, m_R,
\end{eqnarray}
where $G(x)=(x^2\ln x^2)/(x^2-1)$,
while $(m_\nu)_{34} = (m_\nu)_{43} = 0$.
Note that the loop-induced Majorana mass is proportional to $m_R$.

Light neutrinos are subject to the cosmological
upper bound \cite{cosmb} of 50 eV on their masses,
unless they have fast decay channels.
In absence of the latter, we saturate this bound with $m_{\nu_\tau}$,
assuming that the quantity in brackets is of order one,
and that all Yukawa couplings are of order unity
in this model \cite{z8}.
We find that the tree level Majorana mass $m_R$
should be less than a few hundred keV.
This may appear unnatural at first sight.
We will return for discussions later.

The muon neutrino $\nu_\mu$ acquires mass via the exchange
of two $\phi_5^\pm$ bosons (in gauge basis)
with overlapping loops, as shown in Fig. 2.
Because the three left-handed neutrinos
do not have right-handed partners,
one cannot have analogous two-loop nested diagrams
as for charged lepton mass generation \cite{z8}.
Fig. 2 is generated as follows.
{}From nearest neighbor Yukawa couplings,
which follow from the $Z_8$ charge assigments,
the left-handed $\nu_{\mu}$ first couples to
a right-handed $\tau$ via the $f_{23}$ term.
An $m_\tau$ insertion is necessary to
flip the chirality of $\tau$,
before another charged $\phi_5$ scalar emission
flips $\tau_L$ into $N_R$ via the $\tilde f_{34}$ term.
After $m_R$ insertion, the sequence is repeated in reverse order.
In all, two $m_\tau$ insertions are necessary,
but ultimately $m_R$ provides the seed.
It is also clear that only the $\phi_5^\pm$ boson
enters the diagram, hence the process does not
depend on scalar mixing in any crucial way.
The contribution from Fig. 2 is
\begin{eqnarray}
  (m_{\nu})_{22} &=&  -(f_{23}\tilde f_{34})^2\, m_R\, m_\tau^2\,
                   \int {d^4p\over (2\pi)^4}{1\over {p^2-m^2_{\tau}}}\;
                   \left [ {{s^2_+}\over {p^2-m^2_{H_1^+}} }+
                           {{c^2_+}\over {p^2-m^2_{H_2^+}} }
                                             \right ]\times   \nonumber \\
              & & \int {d^4q\over (2\pi)^4}{1\over {q^2-m^2_{\tau}}}\;
                       {1\over {(p+q)^2}}\;
                   \left [ {{s^2_+}\over {q^2-m^2_{H_1^+}} }+
                           {{c^2_+}\over {q^2-m^2_{H_2^+}} }
                                               \right ].
\label{eq:Mnu22}
\end{eqnarray}
In the above $s^2_+$ and $c^2_+$ are short-hand for $\sin^2\theta_+$ and
$\cos^2\theta_+$, respectively, and we have also
ignored $m_R$ in the denominator since $m_R \ll m_\tau$.
This approximation simplifies the calculation.
The presence of two tau mass insertions makes the diagram finite.
Note that the two-loop Dirac mass term
$(m'_{\nu})_{24}\bar \nu_{\mu L} N_R$ can be rotated away
and does not contribute to $m_{\nu_\mu}$ directly,
but modifies the Dirac mass for $N$ slightly.
Since each power of $m_\tau$ is at one-loop order \cite{z8},
Fig. 2 is effectively at four-loop order.
In contrast, the contribution from the exchange
of two W bosons \cite{twow} is effectively at
six-loop order within the model, since mixing in
leptonic charged current is also loop induced \cite{z8}.
Hence, numerically it can be neglected in our model.
%The diagram is analogous to Fig. 2 but with $W$ boson exchange.

Eq. (\ref{eq:Mnu22}) can be reduced to
\begin{eqnarray}
  (m_{\nu})_{22}
         &=& -\left({{f_{23}\tilde f_{34}}\over {16\pi^2}}\right)^2\,
                          m_R\, m_\tau^2\,
                      \sum_{i,j=H_1^+,H_2^+}\;
           {{g^{(\mu)}_{ij}}\over{(m_i^2-m^2_\tau)(m_j^2-m^2_\tau)}}\;
                      \times   \nonumber \\
         & & \int_0^1 dx \{ m_j^2
               [-Li_2(-u)]_{u=(a_{\tau j}-x)/x}^{u=(a_{ij}-x)/x} \;
                          - m_\tau^2
               [-Li_2(-u)]_{u=(1-x)/x}^{u=(a_{i\tau }-x)/x} \},
\label{eq:M22}
%In eq. (\ref{eq:M22}),
%
%         & & \int_0^1 dx [m_j^2\int_{(a_{\tau j}-x)/x}^{(a_{ij}-x)/x}\;
%                                du {{ln(1+u)}\over u}
%                      -m_\tau^2\int_{1-x)/x}^{(a_{i\tau}-x)/x\;
%                                du {{ln(1+u)}\over u} ],
%
\end{eqnarray}
where
\begin{eqnarray}
g^{(\mu)}_{ij} &=& \left(
 \begin{array}{cc}
  s^4_+        &  s^2_+ c^2_+ \\
  s^2_+ c^2_+  &  c^4_+
 \end{array} \right),
%\label{eq:UK}
%In eq. (\ref{eq:UK}),
                   \nonumber \\
           a_{xy} &=& m^2_x/m^2_y,\ \ ({x,y=i,j, \tau}).
\end{eqnarray}
Here, $Li_2(u)$ is the dilogarithm
(polylogarithmic function of order 2).
The integral in eq. (\ref{eq:M22}) is evaluated numerically.
Note that the Majorana mass $(m_\nu)_{24}$ is of
the same loop order as $(m_\nu)_{22}$.
Nevertheless, it is still reasonable to use $(m_\nu)_{22}$
to represent $m_{\nu_\mu}$
since these higher order masses are much less than
the tree level mass $m_R$.
To estimate the range of $m_{\nu_\mu}$,
we will take the maximal value $m_R=0.5$ MeV (the electron mass!)
in eq. (\ref{eq:M22}).
As mentioned in ref. \cite{z8}, the nonstandard Higgs bosons
cannot be too far above the electroweak scale
while all Yukawa couplings should be of order unity.
We therefore take
$f_{23} = \tilde f_{34} = \sqrt{2}$
and fix $\sin\theta_+=0.2$.
We plot in Fig. 3 $m_{\nu_\mu}$ versus
$m_{H_2^+} \in (125,\ 4000)$ GeV,
for $m_{H_1^+} = $ 125, 250, 500, 1000 GeV.
We see that $m_{\nu_\mu}$ can reach up to
the order of $10^{-2}$ eV. It is likely to be within
the range of $10^{-3} - 10^{-2}$ eV for $m_{H_2^+}$
below $2$ TeV. In general, it is insensitive to
$m_{H_1^+}$ if $\sin\theta_+$ is not too large,
since Fig. 2 is due to $\phi_5^\pm$ alone in gauge basis.

The electron neutrino $\nu_e$ acquires mass via the
exchange and {\it mixing} of $\phi_3^-$ and $\phi_5^-$,
as shown in Fig. 4.
The left-handed electron neutrino flips to a
right-handed muon by emitting a $\phi^+_3$ boson
via the $f_{12}$ coupling.
The right-handed muon then emits a $\phi_5^-$ boson
and changes into left-handed tau neutrino.
Since $\nu_\tau$ has acquired Majorana mass via Fig. 1,
$m_{\nu_\tau}$ now effectively functions as seed.
The sequence is then repeated in reverse order.
It is clear that $\phi_3^\pm$--$\phi_5^\pm$ mixing
is needed. Since $m_{\nu_\tau}$ is effectively
at one-loop order, this is
effectively a three-loop diagram.
The contribution from Fig. 4 is
\begin{eqnarray}
  (m_{\nu})_{11} &=& -(f_{12} f_{32})^2\, s^2_+ c^2_+ \,
                                     m_{\nu_\tau}\,
                   \int {d^4p\over (2\pi)^4}{{\not p}\over {p^2}}\;
                   \left [ {{-1}\over {p^2-m^2_{H_1^+}} }+
                           {{1}\over {p^2-m^2_{H_2^+}} }
                                             \right ] \times  \nonumber \\
              & & \int {d^4q\over (2\pi)^4}{{\not q}\over {q^2}}\;
                       {1\over {(p+q)^2}}\;
                   \left [ {{-1}\over {q^2-m^2_{H_1^+}} }+
                           {{1}\over {q^2-m^2_{H_2^+}} }
                                               \right ],
\end{eqnarray}
where we have ignored $m_{\nu_\tau}$ in the denominator
since $m_{\nu_\tau} \ll m_{H_1^+}, m_{H_2^+}$.
After some calculation the above equation
becomes
\begin{eqnarray}
  (m_{\nu})_{11} = \left({{f_{12} f_{32}}\over
                            {16\pi^2}}\right)^2\,
                      s^2_+ c^2_+\, m_{\nu_\tau}\,
                 & &    \sum_{i,j=H_1^+,H_2^+}\;
                   {g^{(e)}_{ij}\, m_j^2\over m_i^2}\;
                                           \times   \nonumber \\
             & &     \int_0^1 x dx\,
              [-Li_2(-u)]_{u=-1}^{u=(a_{ij}-x)/x},
\label{eq:M11}
%
%         & & \int_0^1 dx [m_j^2\int_{(a_{\tau j}-x)/x}^{(a_{ij}-x)/x}\;
%                                du {{ln(1+u)}\over u}
%                      -m_\tau^2\int_{1-x)/x}^{(a_{i\tau}-x)/x}\;
%                                du {{ln(1+u)}\over u} ],
%
\end{eqnarray}
where
\begin{equation}
g^{(e)}_{ij} = \left(
 \begin{array}{rr}
   1  & -1  \\
  -1  &  1
 \end{array} \right).
\end{equation}
The diagram is logarithmically divergent, but the GIM-like mechanism of
$\phi_3^\pm$--$\phi_5^\pm$ mixing renders the contribution finite.
To evaluate the plausible range for $m_{\nu_e}$, we take
$f_{12}= f_{32}=\sqrt 2$, $m_{\nu_\tau} \sim 50$ eV,
and plot, in Fig. 5, $m_{\nu_e}$ versus
$m_{H_2}^+$ with $m_{H_1}^+=$ 125, 250, 500 1000 GeV,
fixing $\sin\theta_+=0.2$.
We see that $m_{\nu_e}$ can reach up to order $10^{-3}$ eV
if $H_1^+$ and $H_2^+$ are far from being degenerate.
However, as expected,
$m_{\nu_e}$ vanishes for $m_{H_1^+} = m_{H_2^+}$
and we see that this GIM-like cancellation effect is rather strong,
and $m_{\nu_e}$ could be far below $10^{-3}$ eV.

Some discussion is in order.
It may be asked why the effectively four-loop
$\nu_\mu$ mass generation diagram is seemingly larger than
the effectively three-loop $\nu_e$ mass generation diagram.
The reason is because Fig. 2 depends on $\phi_5^\pm$ alone,
and is not sensitive to charged scalar mixing.
Indeed, we have taken the mixing angle $\sin\theta_+$ as ``small"
and the result depends largely on $m_{H_2^+}$, the physical
charged scalar that is mostly the $\phi_5^+$ in gauge basis.
In contrast, the mechanism in Fig. 4 for $\nu_e$ mass generation
depends crucially on $\phi_3^\pm$--$\phi_5^\pm$ mixing.
It provides a GIM-like mechanism that not only
leads to divergence cancellation,
it further suppresses the three loop contribution
in an interesting parameter domain.
{}From the prespective of the model,
$H_1^+$ and $H_2^+$ mass should not be too widely apart \cite{z8},
hence for the plausible range of $m_{H_{1,2}^+} \in (0.5,\ 2)$ TeV,
$m_{\nu_e}$ is rather suppressed. But outside of this domain,
if $H_1^+$--$H_2^+$ splitting is rather sizeable, one in general finds
Fig. 4 dominating over Fig. 2, as expected.

Second,
although all light neutrino Majorana masses vanish with $m_R$,
note that $\nu_\mu$ mass is directly related to $m_R$
and suffers from no cancellation mechanism.
In contrast, $m_{\nu_\tau}$ and $m_{\nu_e}$
have separate cancellation mechanisms in the form of scalar mixing,
and in fact  $m_{\nu_e}$ depends on $m_R$ through
an $\nu_\tau$ mass insertion.
The reason behind the difference between $m_{\nu_\mu}$
{\it vs.} $m_{\nu_e}$, $m_{\nu_\tau}$ is the remnant
$Z_2$ symmetry \cite{z8} after soft-breaking of $Z_8$:
$\nu_\mu$ and $N_L$ is even,
while $\nu_e$ and $\nu_\tau$ are odd,
as are the nonstandard scalar bosons.
Thus, $\nu_e$ and $\nu_\tau$ belong to the same class,
while $\nu_\mu$ is more closely related to $N_L$.
This dichotomy of leptons is very useful in suppressing dangerous
tranistions like $\mu\to e\gamma$.
Third,
it is of interest to look into questions related to
neutrino oscillations.
Inspection of Fig. 4 suggests that one has a $\nu_e$--$\nu_\tau$
Majorana mixing term also of three-loop order.
The only difference is that for one of the scalar loops,
one has $\phi_5^\pm$ throughout, hence one is less sensitive
to the GIM like $\phi_3^+$--$\phi_5^+$ mixing effect,
although it is still needed for sake of divergence cancellation.
Note however that this off-diagonal term is still at three-loop
order and proportional to the one-loop $m_{\nu_\tau}$
mass.
Hence, we estimate the $\nu_e$--$\nu_\tau$ mixing angle
to be of order $10^{-5} - 10^{-4}$, which is almost independ of $m_{\nu_\tau}$
value, but depends on the fact that there are two extra loops
compared to Fig. 1.
The mass eigenvalue is barely changed from that of eq. (\ref{eq:M11}).
The mass difference between $\nu_e$ and $\nu_\tau$ is
basically just $m_{\nu_\tau}$.
We have used it to saturate the cosmological bound,
but it can be lowered if one is willing
to entertain a lower $m_R$ value.
However, since the tiny mixing angle is more or less
independent of $m_{\nu_\tau}$, we find that our model
cannot provide a basis for a MSW-type solution to
the solar neutrino problem.
Finally,
we have to face the issue of naturalness of the $m_R$ scale.
One of the most interesting aspects of our radiative model for
charged lepton mass generation \cite{z8}
is that all dimensional parameters are of order weak scale,
while all dimensionless parameters are of order one.
To allow for Majorana neutrino mass generation, however,
we need $m_R$ to be less than one MeV,
which seems to run against the spirit of the model.
We note, however, that $m_R$ in the model is a
very different parameter from all the others.
It is not related in anyway to the Higgs bosons introduced.
It is basically a free parameter in the same way that
$m_e$ is a parameter for QED.
It is clear that setting it to zero one
restores a larger chiral symmetry.
%It also seems to be intimately linked to a fundamental problem of nature,
%the deficiency of solar neutrinos.
We therefore would like to advocate a lenient attitude towards
the naturalness question for $m_R$.

In summary, if new sequential leptons are found at the weak scale,
the traditional seesaw mechanism for neutrino mass generation,
as well as $SO(10)$ based GUT theories, may be in jeopardy.
One may therefore face a serious challenge with
the solar neutrino problem.
We provide a new radiatve mechanism for generating
tiny Majorana neutrino masses.
Because of remnant $Z_2$ symmetry,
$\nu_\tau$ and $\nu_e$ receive
Majorana mass in a different way than $\nu_\mu$,
although in each case the tree level Majorana mass
for heavy neutral lepton, $m_R$, provides the seed.
If one saturates the cosmological bound with $m_{\nu_\tau}$,
then $m_{\nu_\mu}$ and $m_{\nu_e}$ are typically of order
$10^{-2}$ and $10^{-3}$ eV, respectively,
although $m_{\nu_e}$ can be considerably smaller.
Unfortunately, $\nu_e - \nu_\tau$ mixing is loop induced
and is too small to sustain an
MSW solution to the solar neutrino problem.

\acknowledgments
We thank E. Ma %, D. Chang and C.-Q. Geng
for useful discussions.
The work of WSH is supported in part by grants NSC 82-0208-M-002-151
and NSC 83-0208-M-002-023,
and GGW by NSC 83-0208-M-002-025-Y
of the Republic of China.

%

%\vskip -1cm
\figure{Radiative mechanism for $m_{\nu_\tau}$.}

\vskip -1cm
\figure{Radiative mechanisms for $m_{\nu_\mu}$.}

\vskip -1cm
\figure{$m_{\nu_\mu}$ {\it vs.} $m_{H_2}^+$
        for (top to bottom) $m_{H_1}^+ = $ 125, 250, 500 1000 GeV
        at fixed $\sin\theta_+=0.2$.}

\vskip -1cm
\figure{Radiative mechanisms for $m_{\nu_e}$.}

\vskip -1cm
\figure{$m_{\nu_e}$ {\it vs.} $m_{H_2}^+$
        for (top to bottom) $m_{H_1}^+ = $ 125, 250, 500 1000 GeV
        at fixed $\sin\theta_+=0.2$.}
\end{document}